\documentclass[%
 reprint,
 amsmath,amssymb,
 aps,
]{revtex4-2}

\usepackage{graphicx}
\usepackage[utf8]{inputenc}% Include figure files
\usepackage{dcolumn}% Align table columns on decimal point
\usepackage{bm}% bold math
\usepackage{xcolor} % for text colors

\begin{document}

\preprint{APS/123-QED}

\title{A Stochastic Cluster Expansion for Electronic Correlation in Large Systems}% Force line breaks with \\
%\thanks{A footnote to the article title}%

\author{Annabelle Canestraight}
 \affiliation{Department of Chemical Engineering, University of California, Santa Barbara, CA 93106-9510, U.S.A.}
\email{acanestraight@ucsb.edu}%Lines break automatically or can be forced with \\
\author{Anthony J. Dominic III}
\author{Andr\'es Montoya-Castillo}
\affiliation{Department of Chemistry, University of Colorado Boulder, Boulder, CO 80309, USA}
\author{Libor Veis}
\affiliation{Department of Theoretical Chemistry, J. Heyrovský Institute of Physical Chemistry, Czech Academy of Sciences, Prague, Czech Republic}
\author{Vojtech Vlcek}
\affiliation{Materials Department, University of California, Santa Barbara, CA 93106-9510, U.S.A.}
\affiliation{Department of Chemistry and Biochemistry, University of California, Santa Barbara, CA 93106-9510, U.S.A.}

\date{\today}% It is always \today, today,
             %  but any date may be explicitly specified

\begin{abstract}
Accurate many-body treatments of condensed-phase systems are challenging because correlated solvers such as full configuration interaction (FCI) and the density matrix renormalization group (DMRG) scale exponentially with system size. Downfolding and embedding approaches mitigate this cost but typically require prior selection of a correlated subspace, which can be difficult to determine in heterogeneous or extended systems. Here, we introduce a stochastic cluster expansion framework for efficiently recovering the total correlation energy of large systems with near-DMRG accuracy, without the need to select an active space \textit{a priori}. By combining correlation contributions from randomly sampled environment orbitals with an exactly treated subspace of interest, the method reproduces total energies for non-reacting and reactive systems while drastically reducing computational cost. The approach also provides a quantitative diagnostic for molecule–solvent correlation, guiding principled embedding decisions. This framework enables systematically improvable many-body calculations in extended systems, opening the door to high-accuracy studies of chemical processes in condensed phase environments.

\end{abstract}

\maketitle

%\tableofcontents
Although a wide range of solution-phase systems can be treated with mean-field electronic structure methods, many critical phenomena require an accurate description of phenomena in relatively small regions associated with strong interactions. Mean-field approaches often fail precisely where chemical complexity is greatest, such as the transition state of a chemical reaction\cite{adqmmm, Karelina2017SystematicSimulation, VanDerKamp2013CombinedEnzymology,Dittrich2003OnF1-ATPase, Chen2023ElucidatingStructure}. On the other hand, a fully quantum-mechanical treatment of systems of this scale is computationally prohibitive. Predictive simulations thus first simplify the problem by separating out the nuclear degrees of freedom, e.g., via the quantum mechanical/molecular mechanical (QM/MM) methods\cite{adqmmm, Karelina2017SystematicSimulation, VanDerKamp2013CombinedEnzymology}. Further, they utilize a broad class of downfolding and embedding strategies, in which a large system is partitioned into a region of primary chemical interest, treated by (nearly) exact wavefunction or Green's function solvers \cite{Knizia2012DensityTheory, Knizia2013DensityTheory, Wouters2016AChemistry, Pernal2016ReducedFunctional,Inglesfield1981AEmbedding, Inglesfield2015TheStructure, Georges1992HubbardDimensions, Kotliar2006ElectronicTheory, Zgid2011DynamicalPerspective, Chibani2016Self-consistentConcept,DvorakRinke2019(ED),DvorakRinkeGolze2019ED,Pham2020-iuDensityMatrixEmbedding,shee2024staticquantumembeddingscheme,Ma2021EmbeddingcRPA,He2020-raFOEmbedding,CanestraightRenorm2025,Romanova2023,Aryasetiawan2009,MiyakeEffevtiveBand,SakumaWerner2013Dynamical,zheng2017cluster,chang2024downfolding,rossmannek2023quantum,li2024interacting,monino2025projection,scott2021extending,nusspickel2022systematic,nusspickel2022systematic,ferretti2024green, kleiner2025quantummontecarloassessment,mejuto2023efficient,mejuto2024quantum,lee2021efficient,chitra2001effective}, and an environment treated at a lower level of theory\cite{Sun2016QuantumTheories,Muchler2022Static}. Using the atomic configurations generated with QM/MM, these embedding and downfolding methods recover electronic correlation effects within the energetically and spatially bounded region, Frontier Chemical Subspace (FCS). The total energy is thus $E = E_{\rm MF} + \varepsilon_c $, where $E_{\rm MF}$ is the mean-field contribution (e.g., Hartree–Fock), and $\varepsilon_c$ is the correlation energy.

\begin{figure}[h!]
    \centering 
    \includegraphics[width=\linewidth, trim=70 70 70 70, clip]
    {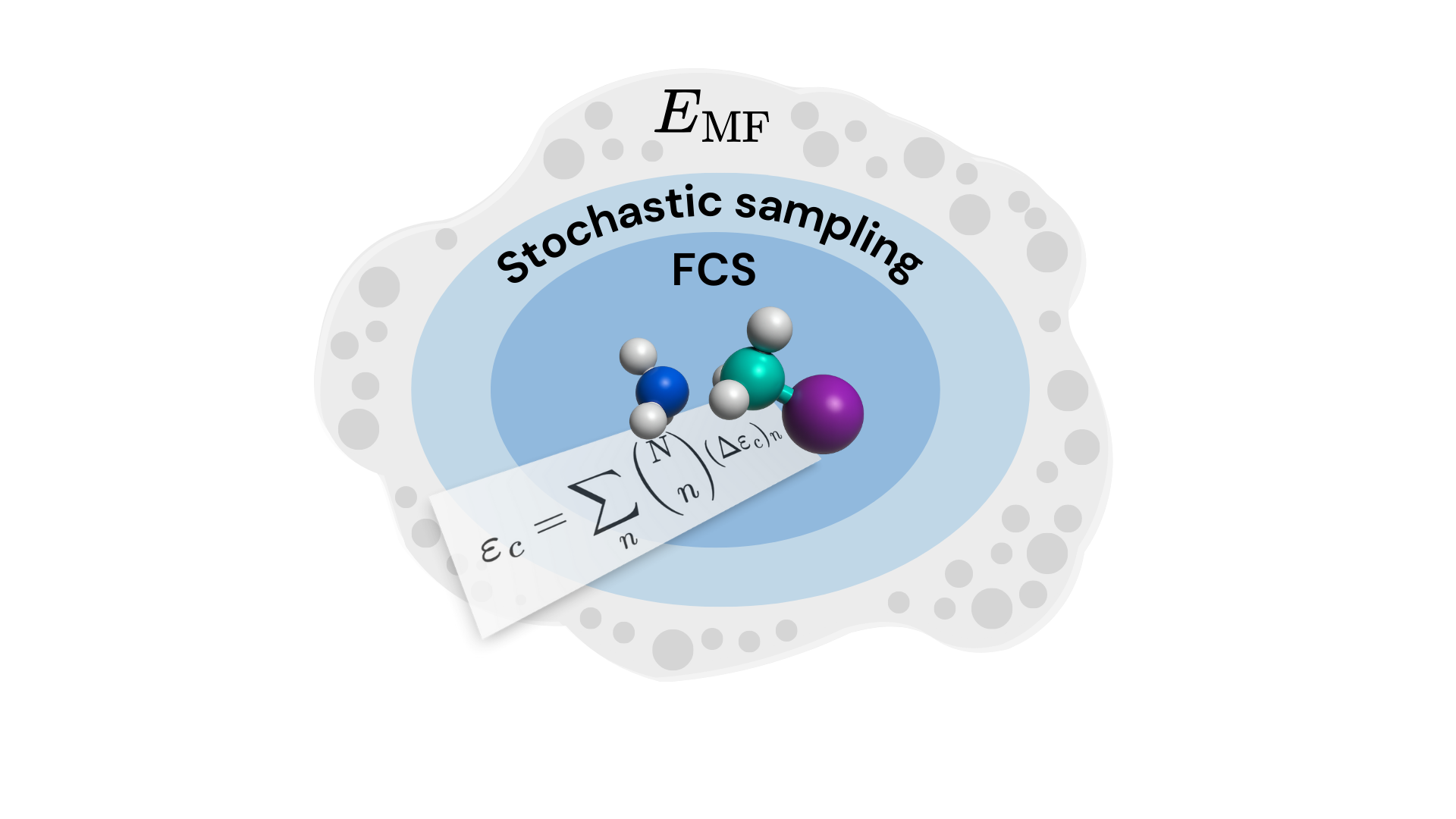}
    \caption{Schematic: the full cluster expansion is used to describe the FCS while a lower order truncation treats the stochastically sampled region around the FCS.}
    \label{fig:fig1}
\end{figure}
A common feature of \textit{ab initio} embedding and downfolding techniques is the need to manually select a partition between the subsystem of interest and its environment\cite{NormSquishActiveSapce}. Related approaches, such as bootstrap embedding and localized active space methods, further subdivide the system into multiple correlated regions \cite{Abraham_2021,TVVBootstrap1,TVVBootstrap2,Hermes_2019, bauman2019downfolding,kowalski2025resource,KowalskiEmbedding}. If strong correlations extend across this boundary, inadequate partitioning can lead to both qualitative and quantitative errors. Importantly, the optimal choice of subspace is often not known \textit{a priori}, particularly for systems undergoing chemical transformation \cite{VanDerKamp2013CombinedEnzymology, Sun2016QuantumTheories,Zlobin2023ChallengesAtoms, Wu2018ParametrizationAtoms,CanestraightRenorm2025}. Consequently, the choice of FCS typically relies on chemical intuition about the coupling between the fragment and environment. Ensuring that all relevant strong-correlation effects are captured further requires systematic convergence with respect to the size of the FCS. In practice, such convergence is difficult to achieve: enlarging the FCS increases the number of electrons treated at a high level of theory, resulting in large computational costs. This lack of a systematic and scalable procedure for defining and converging the correlated subspace represents a major hurdle for embedding and downfolding approaches.

In this Letter, we overcome these limitations by introducing a new framework for computing the otherwise intractable environmental contributions to electronic correlation in complex nanoscale systems. The method utilizes randomized sampling of single-particle orbitals to compress the problem size, which substantially reduces the sensitivity to the choice of partitioning while accurately yielding the total correlation energy $\varepsilon_c$. Further, the methodology substantially reduces the computational cost by effectively sampling the correlation terms and avoiding the treatment of large Hilbert spaces. We illustrate that the approach systematically approaches the numerically exact total correlation energy. A distinct advantage of our approach is its use of a cluster expansion of the total correlation energy, in which contributions arising from all groupings of single-particle states are formally organized and systematically evaluated without the need for chemical intuition. Specifically,
\begin{equation}
         \label{eq:clusterexpansion}
         \varepsilon_c =  \sum_n \binom{N}{n} (\Delta\varepsilon_{c})_n,
    \end{equation}
where $(\Delta\varepsilon_{c})_n$ is the contribution to the total correlation energy arising from each $n$-body combination of single-particle orbitals. Cluster expansions of this form are widely used in many contexts, including energy expansions in alloys and molecular systems, and can be shown to recover the exact correlation energy when evaluated to all orders \cite{PaesaniCluster1,PaesaniCluster2,Nesbet1968}.
   
Exact evaluation of Eq.~\eqref{eq:clusterexpansion} for large systems is computationally infeasible, as computing the $N$th-order term is equivalent to solving the full many-body problem. However, truncating the expansion to finite order may cause higher-order correlation effects to be neglected within the chemically active subspace, where strong correlations may be essential for an accurate description. In this work, we consider a rather common scenario in which the system (composed of occupied and virtual single-particle orbitals) can be separated into two sets of single-particle orbitals based on their level of correlation; the subsystem of interest (FCS) is selected based on both locality and proximity to the Fermi level. The FCS is treated without truncation (i.e., numerically exactly), while the expansion is truncated for the environmental space. The resulting correlation energy cluster expansion for the partitioned system can be written as follows,
    \begin{equation}
\label{eq:splitcluster}
        \varepsilon_c \approx \varepsilon_c^{\rm FCS} + \sum_\phi \Delta\varepsilon_c^\phi +\sum_{\phi \phi'} \Delta\varepsilon_c^{\phi \phi'} + \cdots.
    \end{equation}
Here, $\varepsilon_c^{\rm FCS}$ denotes the correlation energy obtained by treating the FCS alone, while the remaining terms account for correlation contributions arising from single-particle states in the environment. The environment cluster terms, $\Delta\varepsilon_c^{\phi}$ and $\Delta\varepsilon_c^{\phi \phi'}$, are defined as the change in the total correlation energy resulting from the inclusion of one or two environment single-particle states, $\phi$ and $\phi'$, into the active space calculation.
The one-body environment contribution is given by
  \begin{equation}
        \label{eq:one-body}
        \Delta\varepsilon^{\phi} = \varepsilon_c^{\rm FCS + \phi} -\varepsilon_c^{\rm FCS}. 
    \end{equation}
which is obtained by evaluating the correlation energy for two active spaces that differ only by the inclusion of orbital $\phi$. An analogous construction yields the two-body contribution, which captures correlation effects between pairs of environment orbitals added jointly to the FCS:
 \begin{equation}
 \label{eq:two-body}
        \Delta\varepsilon^{\phi \phi'} = \varepsilon_c^{\rm FCS + \phi + \phi'} -\varepsilon_c^{\rm FCS} - \Delta\varepsilon^{\phi}- \Delta\varepsilon^{\phi'}.
    \end{equation}

Evaluating the cluster expansion terms in this manner ensures that correlations between the FCS and the surrounding environment are treated using the same underlying many-body theory employed for the FCS itself. Moreover, the resulting expression for the total correlation energy is systematically improvable: higher-order terms may be included as needed. For the systems examined in this Letter, however, truncation at second order is found to reproduce the total correlation energy.

While using the truncated cluster representation (Eq.~\ref{eq:splitcluster}) already greatly reduces the dimension of the Hamiltonians that must be solved, the \emph{number} of Hamiltonians that must be solved can be greatly reduced through a stochastic formulation of the cluster expansion. In this approach, single-particle orbitals spanning the environment space are sampled to construct a stochastic orbital,
\begin{equation}
        |\zeta\rangle =\frac{1}{\sqrt{N_{\rm R}}} \sum_j^{N_{\rm R}} e^{i\theta_j} |\phi_j\rangle,
    \end{equation}
where $\theta_j$ is a random phase chosen uniformly from the interval $[0,2\pi]$, $|\phi_j\rangle$ is a single-particle orbital from the rest space, and $N_R$ is the number of states in the rest space. For many-body solvers restricted to real-valued orbitals, the phase factor may be confined to the real axis by choosing $\theta_j = 0$ or $\pi$. The correlation contributions associated with stochastic orbitals $\zeta$ (and $\zeta'$) are evaluated using Eqs.~\ref{eq:one-body} and \ref{eq:two-body}. Because each stochastic orbital represents an evenly weighted sampling of all environment single-particle states, this procedure yields the average correlation contribution of a single environment orbital, or of a pair of environment orbitals, to the FCS. In systems where the environment contains significant redundancy, such as many chemically equivalent solvent molecules, the number of explicit many-body calculations required to estimate the total correlation energy can therefore be dramatically reduced.

These averaged contributions are weighted by the number of single-particle states and state pairs they represent to obtain an expectation value for the total correlation energy,
  \begin{equation}
        \langle\varepsilon_c \rangle \approx  \varepsilon_c^{\rm FCS} + \langle N_{\rm R} \Delta \varepsilon_c^{\zeta} + \frac{N_{\rm R}(N_{\rm R}-1)}{2} \Delta\varepsilon_c^{\zeta \zeta'}\rangle_{N_{\zeta}}.
    \end{equation}
This expression is exact in the limit of infinite sampling. However, for a finite number of samples, the stochastic error associated with this expected value decays as $1/\sqrt{N_{\zeta}}$, where $N_{\zeta}$ denotes the number of independent stochastic samples. We will show this is indeed the asymptotic behavior of the computational error of the stochastic cluster expansion (SCE). Similar sampling strategies have been employed in a variety of contexts to exploit redundancy in large environments and have been shown to substantially reduce overall computational cost \cite{Vlcek2017-rz,Vlcek2018Swift,Canestraight2024Efficient,RoiDannyGW,annurev:/content/journals/10.1146/annurev-physchem-090519-045916,gao2015sublinear,neuhauser2013expeditious,rabani2015time}. While many repeat calculations are required to converge this expectation value of the correlation energy, this is still computationally cheaper than a single calculation of the full space for a large system.

An additional advantage of the present formulation is its compatibility with a wide range of many-body solvers. Systematically improvable wave-function-based methods, including Møller–Plesset perturbation theory (MP2) \cite{Muller1934,neuhauser2013expeditious2,ge2014guided}, and configuration interaction (CI) \cite{DavidSherrill1999}, may be employed to evaluate individual cluster contributions. The results presented in this Letter were obtained using the quantum chemical density matrix renormalization group (DMRG) as the many-body solver \cite{DMRG_White, White1999, chan_review, reiher_perspective, MOLMPS}.

Treating both intra-FCS correlation and FCS–environment inter-correlation using the same many-body solver removes any potential inconsistencies and errors due to double counting. The choice of FCS determines the computational cost; hence, it is advantageous to select the smallest FCS capable of capturing all strong correlation\cite{Golub2021}. For a small FCS, the cost of evaluating the terms of the cluster expansion will be tractable, despite the need to repeat the calculation multiple times. 

To examine the proposed method, we consider a non-reacting model system consisting of sodium metaphosphate in water---a system of biological relevance coupled to a polar solvent. The cluster expansion is evaluated explicitly for the metaphosphate molecule together with its nearest seven water molecules, forming a minimal model system that is exactly solvable using DMRG. The remaining solvent molecules are treated as a mean-field external potential \emph{including} both Hartree and exchange interactions with the environment orbitals. Details of system preparation are found in the SI. We vary the size of the occupied portion of the FCS, holding the number of virtual orbitals constant.

In Fig.~\ref{fig:fig1}, we observe that as the FCS size is reduced, the formally exact contribution of the FCS to the total correlation energy decreases, as expected. Yet, the total correlation energy predicted by the stochastic cluster expansion remains nearly constant across all partitionings. This value is consistently within the standard error of the mean (SEM) of the exact DMRG solution for the full system, indicating that the stochastic cluster expansion is capable of reproducing DMRG-level results independent of the choice of FCS. This is true for even a very small number of samples, indicating that the method doesn't suffer from bias. 

   \begin{figure}
        \centering 
        \includegraphics[width=\linewidth]
        {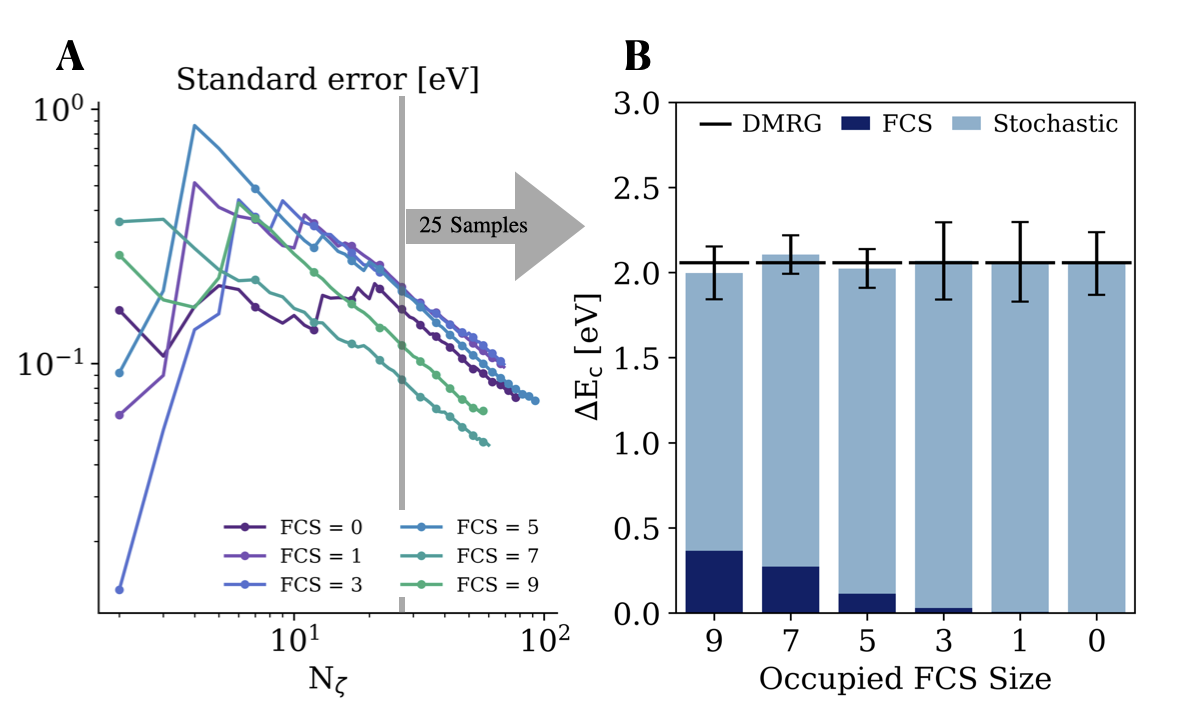}
        \caption{The stochastic cluster expansion method captures the total correlation energy independently of the FCS partitioning of the solvated sodium metaphosphate system. (A) For each FCS size in panel B, the stochastic convergence scales as $1/\sqrt{N_{\zeta}}$. (B) For 25 samples, the total correlation energy for Metaphosphate and 7 waters is repeated for 5 different FCS sizes. We include the same 10 unoccupied single-particle states in all FCS subspaces but add progressively lower energy occupied single-particle states.}
        \label{fig:fig1}
    \end{figure}

Across all FCS–environment partitionings, the SEM decreases as $1/\sqrt{N_{\zeta}}$ within approximately $25$ samples, demonstrating that the number of stochastic samples required to reach a target accuracy can be reliably extrapolated. The error associated with a given partitioning depends on many factors. Most importantly, the stochastic terms are multiplied by both one- and two-body prefactors, such that the error per sample grows up to quadratically with increasing system size. The rate of this growth naturally depends on whether one- or two-body terms dominate. In addition, the stochastic error increases with the magnitude of the environment’s per-electron contribution to the total correlation energy. In the limiting case of a completely uncorrelated FCS and environment, this contribution vanishes, and no stochastic error is introduced. For the results shown in Fig.~\ref{fig:fig1}, the total error for the $80$-electron system ($8$ molecules) falls below $100$~meV for all partitionings within $100$ samples, with additional sampling required to achieve chemical accuracy for the full solute–solvent cluster.
For all partitionings considered, even in this fully solvable system, a computational speedup is achieved for a target accuracy of $100$~meV. For an FCS comprising three occupied and ten unoccupied orbitals, a single stochastic sample requires only $0.18\%$ of the CPU time of the exact DMRG calculation, corresponding to an $86\%$ reduction in computational cost at $100$~meV error. For larger systems that cannot be solved exactly, the relative computational savings are expected to increase further, as fully correlated calculations may be computationally prohibitive.

\begin{figure}
\centering
\includegraphics[width=\linewidth]{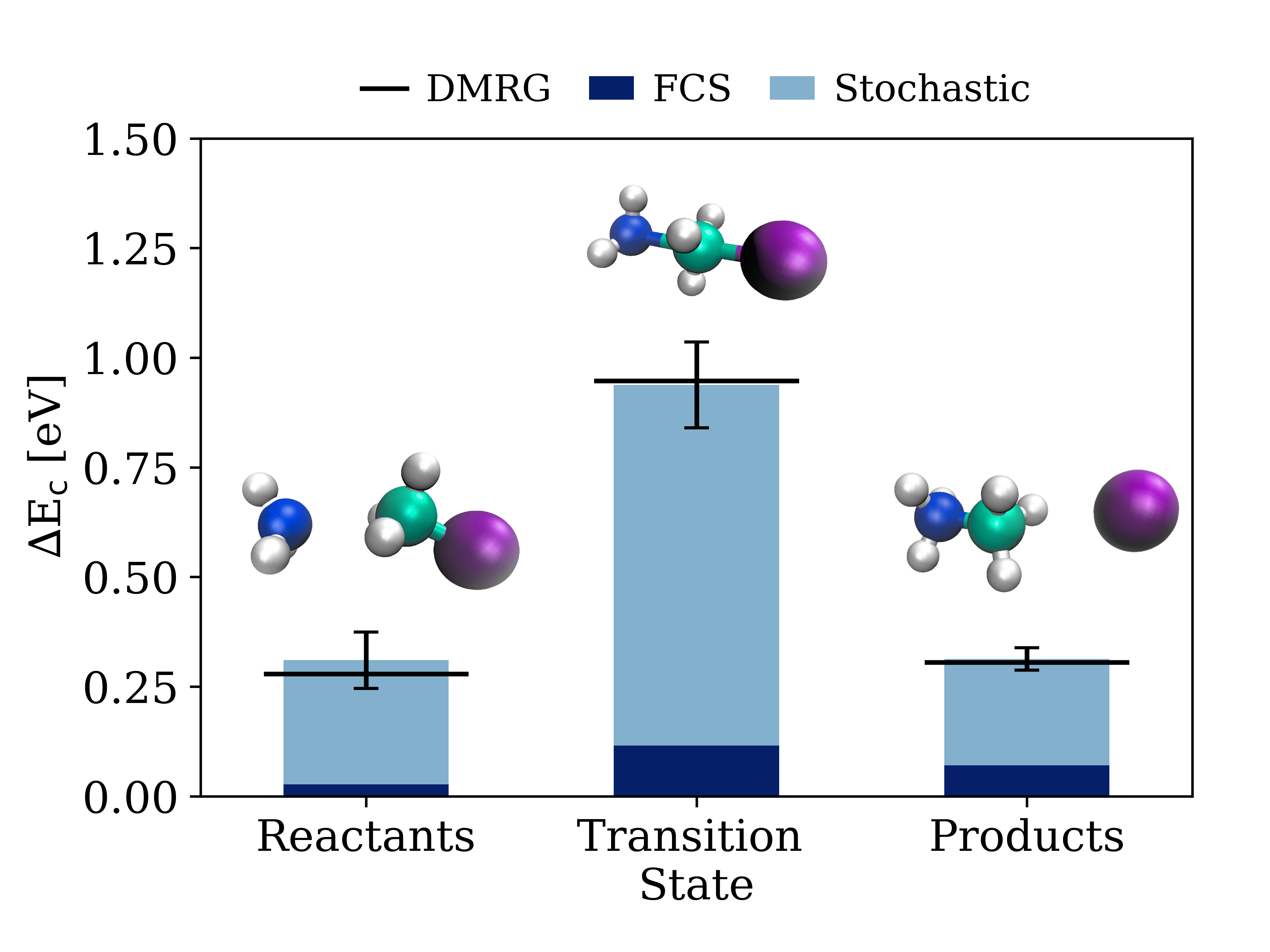}
\caption{The stochastic cluster expansion approach captures the total correlation energy for the Menshutkin reaction $
\mathrm{H_3N + CH_3Cl \;\rightarrow\; CH_3NH_3^{+} + Cl^{-}}$. For an FCS subspace of five occupied orbitals and all eight bound unoccupied orbitals, the total correlation energy is computed for the reactants, transition state, and products of the reaction of $\rm CH_3Cl$ and $\rm NH_3$. Error bars indicate the SEM over 25 stochastic samples.}
\label{fig:fig2}
\end{figure}

The SCE approach performs similarly well for chemically reactive systems, such as transition states along a reaction pathway, where one expects larger contributions to the correlation energy. We compute the total correlation energy for the reaction of $\rm CH_3Cl$ with $\rm NH_3$ (the Menshutkin reaction) solvated by the nearest five water molecules for three representative configurations: reactants, transition state, and products. Details of system preparation are found in the SI. This simple and well-studied reaction serves as a means to benchmark the methodology on a more realistic bond-breaking/formation event, which one would predict, \textit{a priori}, to have a higher correlation energy than the product or reactant states. Figure ~\ref{fig:fig2} demonstrates that the SCE successfully captures the enhanced correlation energy at the substantially more correlated transition state, where the $\rm N-Cl$ bond is breaking and the $\rm N-C$ bond is forming.

 The five-water solvent shell used here serves as a minimal model to capture the transition state correlation energy, rather than as a fully converged description of the solvation environment. In general, the number of solvent molecules necessary to converge reaction energetics depends on the specific reaction. The solvent effects often play a critical role in the reaction mechanism, and the correlation energy of these contributing degrees of freedom must be accounted for in the subsystem's total energy. This correlation is known to decay for solvent electrons that are spatially distant from the subsystem \cite{GwenSpatialDecay}.
The two-body inter-correlation term, $\Delta\varepsilon_c^{\zeta \zeta'}$ (Eq.~\ref{eq:two-body}), allows direct assessment of correlation between orbitals in different spatial regions of the simulation cell. This is achieved by repeatedly sampling two stochastic orbitals, $\zeta$ and $\zeta'$, from different subsets of single-particle states rather than sampling both from the entire environment, as in the previous benchmarking. The expectation value of the two-body correlation energy between $\zeta$ and $\zeta'$ provides direct information about how electronic correlation decays in real space.

To demonstrate this capability, we return to the fully solvated metaphosphate system, this time accounting for the nearest 27 water molecules with the SCE. Using the Pipek–Mezey Wannierization scheme\cite{GwenReducedScaling,GwenEfficientTreatment}, we construct multiple shells of charge density on water molecules selected according to their mean distance from the metaphosphate. The FCS in this calculation includes only a subset of virtual orbitals. Stochastic orbital $\zeta$ is sampled from single-particle states localized on the metaphosphate, while $\zeta'$ is sampled from one of the solvent shells.

\begin{figure}
\centering
\includegraphics[width=1.2\linewidth]{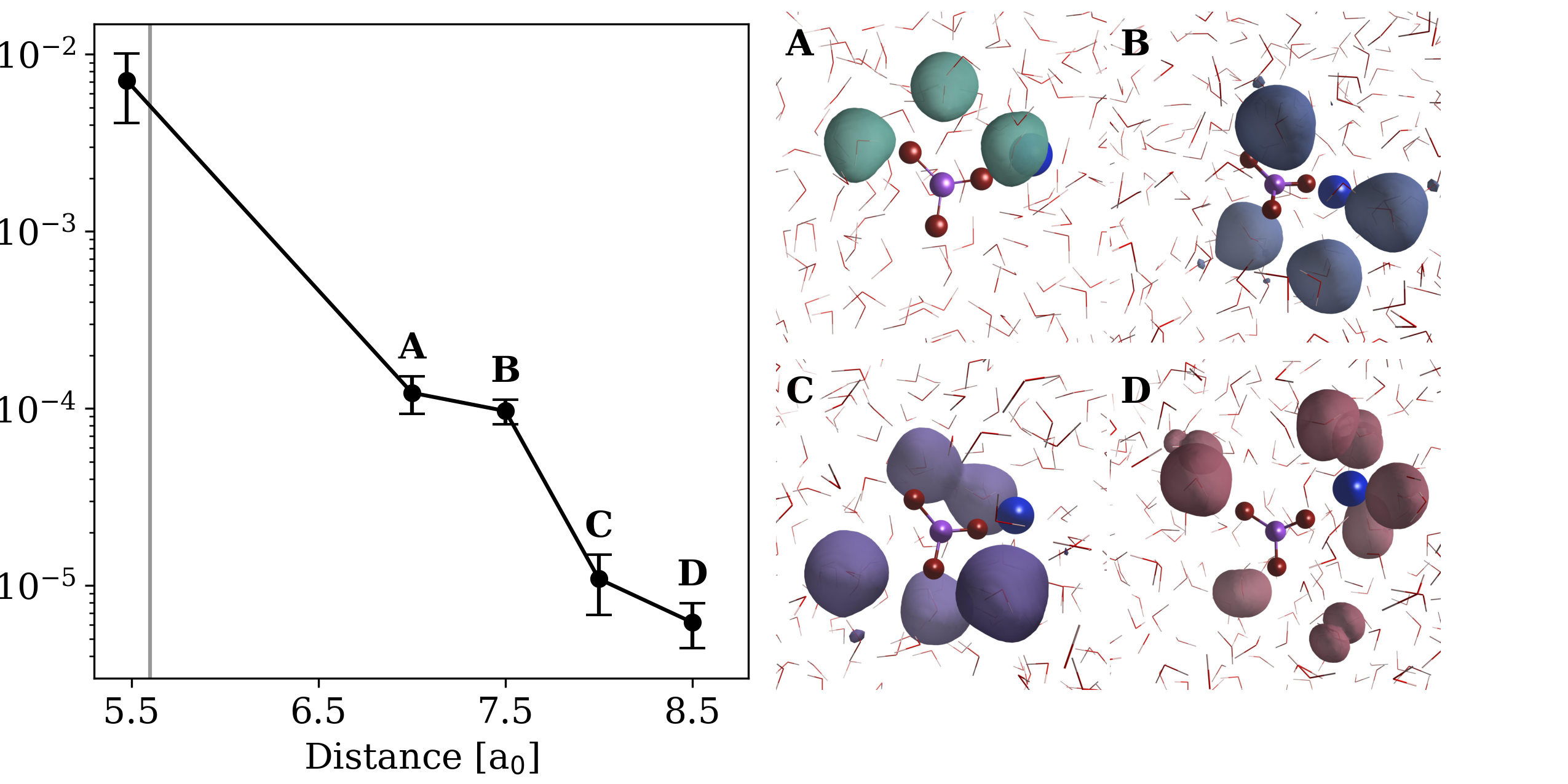}
\caption{(Left) Two-body correlation terms for single-particle states localized on solvent molecules at varying distances from the metaphosphate. The distance in Bohr corresponds to the cutoff used to select water molecules. Symmetric error bars indicate the SEM over 25 stochastic samples. (Right) Panels A–D show the density of single-particle states sampled for each point in the left-hand plot.}
\label{fig:solvent}
\end{figure}

In Fig.~\ref{fig:solvent}, we observe that for two single-particle states drawn from the metaphosphate (below a distance of 5.475 bohr), the two-body correlation term is $\Delta\varepsilon_c^{\zeta\zeta'} = 0.0072 \pm 0.0040~\rm eV$. For the nearest three water molecules, located approximately 7 Bohr from the metaphosphate, the two-body interaction term decreases by nearly two orders of magnitude, indicating substantially reduced correlation between the molecule and even the closest solvent molecules.

This observation suggests that the solvent’s effect on non-reacting water-solvated metaphosphate is well described by a mean-field external potential. Any increase in the total correlation energy upon adding more water molecules to the subsystem arises primarily from the extensivity of correlation energy, rather than from significant correlation between the molecule and solvent.

Beyond computational affordability, the method offers an efficient quantitative diagnostic for monitoring the evolution of molecule–solvent correlation along a reaction trajectory, a capability largely absent from existing methodologies. This provides a principled means of assessing ``solvent (non) innocence,'' ensuring that relevant electronic degrees of freedom are included when necessary, while avoiding costly enlargement of the correlated subspace when they are not. 

The method has been benchmarked for both an ionic compound in a polar solvent and for a reactive bond-breaking event, demonstrating robust performance across chemically distinct regimes. In the limit of many stochastic samplings, it has been shown to systematically approach DMRG-level accuracy, while offering substantial gains in computational efficiency. This efficiency arises from the tradeoff between the cost of repeat solutions of small Hamiltonians defined on a reduced correlated subspace, rather than the direct treatment of a prohibitively large Hamiltonian describing the full system. A definitive advantage is obtained when the number of environment orbitals exceeds the number of stochastic samples required to reach a target accuracy, a condition that is commonly satisfied in condensed-phase systems with significant redundancy, such as environments composed of many identical solvent molecules. The resulting partitioning is therefore most advantageous for very large systems (100's of electrons or greater), where even approximate many-body methods become intractable. In this regime, the approach enables the application of high-accuracy solvers such as DMRG to problems that would otherwise be accessible only at the mean-field level, substantially extending the reach of correlated electronic structure methods.

A current limitation of the approach is that it provides direct access only to the total correlation energy, rather than an explicit many-body wave function. While this restricts the direct evaluation of wave-function–dependent observables, it is consistent with the primary objective of efficiently recovering accurate reaction energetics in large, condensed-phase systems, where total energies and energy differences are often the quantities of greatest interest.

The present formulation of the approach samples correlations at the level of individual orbitals and small orbital groupings, efficiently capturing local renormalization effects. However, it does not explicitly resolve collective, resonant modes involving coherent correlations across large portions of the environment. This can lead to slow or non-systematic convergence with respect to cluster size when delocalized correlations dominate. This behavior reflects the choice of single-particle basis rather than a fundamental limitation of the cluster-expansion framework. Reformulating the expansion in terms of appropriate collective modes would capture such correlations at lower order. In this sense, the convergence properties of the cluster expansion indicate whether a local-orbital or collective-mode description is most appropriate. This direction will be pursued in future work.

The flexibility of the formulation further enhances its applicability. Because the expansion is agnostic to the choice of many-body solver, it can be combined with a wide range of electronic structure methods, from exact diagonalization and coupled cluster to perturbative approaches, enabling high-accuracy studies of chemical reactions in complex, condensed-phase environments at a fraction of the cost of more traditional correlated techniques.

While we have demonstrated that our current implementation of the SCE captures ground-state properties, a natural next step is to extend our framework to excited states. More broadly, the stochastic and fragment-based character of the approach suggests potential relevance for quantum computing implementations of electronic structure, where reconstructing global observables from repeated small calculations leads to resource-lean algorithms and their implementation.

\section*{Acknowledgments}
  The authors would like to thank Brenda Rubenstein, Grant Rotskoff, and Norman Tubman for insightful discussions. The theoretical development and numerical implementation (A.C., L.V. and V.V.), and the QM/MM simulations (A.J.D. and A.M.C.) were supported by Wellcome Leap as part of the Quantum for Bio Program. A.J.D. also acknowledges support from the NIH/CU Molecular Biophysics Program and the NIH Biophysics Training Grant T32 GM145437. L.V. acknowledges support from the Czech Science Foundation (Grant No.~25-18486S).

\bibliography{bib,ajdrefs}
\end{document}